\documentclass[aps,prb,amsmath,twocolumn]{revtex4}
\usepackage{bm}
\usepackage{graphicx}

\begin{document}
\title{Non-local electron transport and
cross-resistance peak in NSN heterostructures}
\author{Mikhail S. Kalenkov}
\affiliation{I.E. Tamm Department of Theoretical Physics, P.N.
Lebedev Physics Institute, 119991 Moscow, Russia}
\author{Andrei D. Zaikin}
\affiliation{Forschungszentrum Karlsruhe, Institut f\"ur Nanotechnologie,
76021, Karlsruhe, Germany}
\affiliation{I.E. Tamm Department of Theoretical Physics, P.N.
Lebedev Physics Institute, 119991 Moscow, Russia}

\begin{abstract}
We develop a microscopic theory describing the peak in
the temperature dependence of the non-local resistance of
three-terminal NSN devices. This peak emerges at sufficiently high
temperatures as a result of a competition between
quasiparticle/charge imbalance and subgap (Andreev) contributions
to the conductance matrix. Both the height and the shape of this
peak demonstrate the power law dependence on the superconductor
thickness $L$ in contrast to the zero-temperature non-local
resistance which decays (roughly) exponentially with increasing
$L$. A similar behavior was observed in recent experiments.
\end{abstract}


\maketitle

Non-local effects in coherent electron transport across hybrid
structures composed of a superconductor (S) attached to two normal
terminals (N) have recently become a subject of intensive
experimental \cite{Beckmann,Teun,Venkat} and theoretical
\cite{FFH,MF,BG,Belzig,KZ06,LY,Duhot,GZ07,KZ07} investigations.
Provided the distance $L$ between two N-terminals (see Fig. 1) is
smaller than (or comparable with) the superconducting coherence
length $\xi$, two {\it non-local} processes contribute to electron
transport through such NSN devices. One process corresponds to
direct electron transfer (DET) between two N-metals through a
superconductor. Another process is the so-called crossed Andreev
reflection (CAR): An electron penetrating into the superconductor
from the first N-terminal forms a Cooper pair together with
another electron from the second N-terminal in which case a hole
goes into the second N-metal. A non-trivial interplay between DET
and CAR yields a rich variety of features observed in recent
experiments \cite{Beckmann,Teun,Venkat}.

Here we focus our attention only on one of such features, a
pronounced peak in the temperature dependence of the non-local
resistance observed in three-terminal NSN structures
\cite{Beckmann,Venkat} and attributed to charge imbalance effects.
Very recently Golubev and one of the authors \cite{GZ07} offered a
theory for this phenomenon interpreting the non-local resistance
peak as a result of a competition between the contributions of
charge imbalance and Andreev reflection. A striking experimental
observation \cite{Venkat} is that the height of the resistance
peak depends on the distance $L$ between N-terminals {\it much
weaker} than the corresponding low temperature cross-resistance
which was found to decay (approximately) exponentially
\cite{Beckmann,Venkat} $\propto \exp (-L/\xi )$ in agreement with
theoretical predictions \cite{FFH,MF,BG,KZ06,KZ07}. Note that due
to the restriction $L \lesssim \xi$ it was not possible to address
the length dependence of the resistance peak
\cite{Beckmann,Venkat} within the model \cite{GZ07}.

Below we will employ the model of three-terminal NSN structures
with ballistic electrodes \cite{KZ06} which allows for a complete
non-perturbative solution of the problem for all values of $L$. We
will specifically address the temperature dependence of the
non-local resistance $R_{12}(T)$ and demonstrate that the height
of the charge imbalance peak scales with $L$ exactly as the
corresponding normal state resistance $R_{N_{12}}$. We believe
that this observation might help to account for recent
experimental findings \cite{Venkat}. In addition we will argue
that -- within the model studied here -- the charge imbalance peak
for $R_{12}(T)$ can occur only in the case of weakly transmitting
NS interfaces and it quickly disappears as the interface
transmissions increase beyond the tunneling limit.

\begin{figure}
\centerline{\includegraphics[width=65mm]{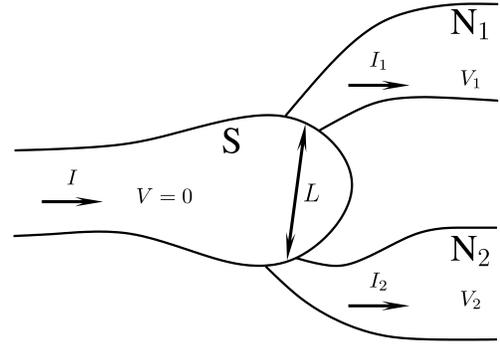}}
\caption{Schematics of our NSN device.} \label{nsn}
\end{figure}

The NSN structure under consideration is depicted in Fig.
\ref{nsn}. We will assume that all electrodes are ballistic
and that both NS interfaces (with cross-sections $\mathcal{A}_1$ and
$\mathcal{A}_2$) have arbitrary transmissions $D_1$ and $D_2$ ranging
from zero to one. The distance between these interfaces $L$ as well
as other geometric parameters are assumed to be much larger than
$\sqrt{\mathcal{A}_{1,2}}$, i.e. effectively both contacts are
metallic constrictions. At the same time the number of conducting
channels ${\cal N}_{1,2}=p_F^2{\cal A}_{1,2}/4\pi$ in each contact
is assumed to be large.

For convenience, we will set the electric potential of the
S-electrode equal to zero, $V=0$. In the presence of bias voltages
$V_1$ and $V_2$ applied to two normal electrodes (see Fig.
\ref{nsn}) the currents $I_1$ and $I_2$ will flow through SN$_1$
and SN$_2$ interfaces. These currents can be evaluated with the
aid of the quasiclassical formalism of nonequilibrium
Green-Eilenberger-Keldysh functions \cite{BWBSZ} $\hat g^{R,A,K}$.
For ballistic electrodes considered here the Eilenberger equations
take the form
\begin{multline}
\left[
\varepsilon \hat\tau_3+
eV(\bm{r},t)-
\hat\Delta(\bm{r},t),
\hat g^{R,A,K} (\bm{p}_F, \varepsilon, \bm{r},t)
\right]
+\\+
i\bm{v}_F \nabla \hat g^{R,A,K} (\bm{p}_F, \varepsilon, \bm{r},t) =0,
\label{Eil}
\end{multline}
where $[\hat a, \hat b]= \hat a\hat b - \hat b \hat
a$, $\varepsilon$ is the quasiparticle energy, $\bm{p}_F=m\bm{v}_F$ is the
electron Fermi momentum vector and $\hat\tau_3$ is the Pauli matrix.
The functions  $\hat g^{R,A,K}$ also obey the normalization conditions
$(\hat g^R)^2=(\hat g^A)^2=1$ and $\hat g^R \hat g^K + \hat g^K \hat g^A =0$.
Here and below the product of matrices is defined as time convolution.

The matrices $\hat g^{R,A,K}$ and $\hat\Delta$ are $2\times2$
matrices in the Nambu space
\begin{equation}
        \hat g^{R,A,K} =
        \begin{pmatrix}
                g^{R,A,K} & f^{R,A,K} \\
                \tilde f^{R,A,K} & \tilde g^{R,A,K} \\
        \end{pmatrix}, \quad
        \hat\Delta=
        \begin{pmatrix}
                0 & \Delta  \\
                -\Delta^* & 0 \\
        \end{pmatrix},
\end{equation}
and $\Delta$ is the BCS order parameter. Without loss of
generality below we choose $\Delta$ to be real. The current
density is related to the Keldysh function $\hat g^K$ by the
standard formula
\begin{equation}
\bm{j}(\bm{r}, t)= -\dfrac{e N_0}{4} \int d \varepsilon
\left< \bm{v}_F \mathrm{Sp} [\hat \tau_3 \hat g^K(\bm{p}_F,
\varepsilon, \bm{r},t)] \right>,
\label{current}
\end{equation}
where $N_0=mp_F/2\pi^2$ is the density of state at the Fermi level and
angular brackets $\left< ... \right>$ denote averaging over the Fermi momentum.

The above equations should be supplemented by the boundary
conditions describing electron scattering at NS interfaces.
Assuming specular reflection at both SN$_1$ and SN$_2$ interfaces
we introduce their transmission probabilities
$D_{1,2}(p_{x_{1,2}})\equiv 1-R_{1,2}(p_{x_{1,2}})$ (where
$p_{x_1}$ ($p_{x_2}$) is the component of $\bm{p}_F$ normal to
SN$_1$ (SN$_2$) interface) and employ the standard Zaitsev
boundary conditions \cite{Zaitsev} in order to match
quasiclassical Green functions at both sides of each of the two
interfaces. Deep inside metallic electrodes S, N$_1$ and N$_2$ the
Green functions should approach their equilibrium values $\hat
g^{R,A}=\pm (\varepsilon\hat\tau_3-\hat\Delta) /\Omega^{R,A}$ in a
superconductor and $\hat g^{R,A}=\pm \hat\tau_3$ in normal metals,
$\Omega^{R,A} =\sqrt{(\varepsilon \pm i\delta)^2-\Delta^2}$. For
the Keldysh functions far from interfaces we have $\hat
g^{K}=\tanh [(\varepsilon +eV\hat\tau_3)/2T](\hat g^{R}-\hat g^{A})$, where
$V=0$, $V_1$ and $V_2$ respectively in S, N$_1$ and N$_2$
electrodes.

The general solution of the problem within the above formalism was
described in details in Ref. \cite{KZ06}. Here we only point out
that the accuracy of the above formalism in the case of
double-barrier structures under consideration is justified
simultaneously by the two conditions \cite{KZ06,GZ02}: ${\cal
A}_{1,2} \ll L^2$ and ${\cal N}_{1,2} \gg 1$.

At low voltages $eV_{1,2} \ll T_c$ we obtain
\begin{gather}
I_1= G_{11}(T) V_1 -G_{12}(T) V_2 ,
\\
I_2=-G_{21}(T) V_1 +G_{22}(T) V_2 .
\end{gather}
where $G_{12}(T)$ and $G_{21}(T)$ are the non-local conductances
of our NSN device \cite{KZ06}:
\begin{multline}
G_{12}(T)=G_{21}(T)=\frac{G_{N_{12}}}{4T}\int \dfrac{d
\varepsilon}{\cosh^2(\varepsilon/2T)}
\\\times (1-{\cal R}_1 |a|^2
)(1-{\cal R}_2 |a|^2 ) \dfrac{1-\tanh^2iL\Omega/v_F}{P({\cal R}_1,
{\cal R}_2)}. \label{Gnonloc}
\end{multline}
Here we defined ${\cal D}_{1,2}\equiv 1-{\cal R}_{1,2}=D_{1,2}(p_F\gamma
_{1,2})$ and $p_F\gamma _{1(2)}$ is normal to the first (second)
interface component of the Fermi momentum for electrons
propagating straight between the interfaces, $\Omega=
\sqrt{\varepsilon^2-\Delta^2}$, $ P(R_1, R_2)=|1-R_1 R_2 a^2 -
Q[\varepsilon (1+R_1 R_2 a^2) + \Delta a(R_1 + R_2)]|^2$,
$Q=\Omega^{-1}\tanh iL\Omega/v_F$, $a=(\Omega-\varepsilon) /\Delta
$,
\begin{equation}
G_{N_{12}}=\frac{8\gamma_1 \gamma_2{\cal N}_1{\cal N}_2{\cal
D}_1{\cal
    D}_2}{R_qp_F^2L^2}
\label{GN12}
\end{equation}
is the non-local conductance in the normal state, $R_q=2\pi/e^2$ is the
quantum resistance unit.

The conductance $G_{11}(T)$ of the first interface
is dominated by the standard BTK expression \cite{BTK}
\begin{equation}
G_{11}(T)= \dfrac{\mathcal{N}_1}{R_q T} \int \dfrac{(1+|a|^2) d
\varepsilon}{\cosh^2(\varepsilon/2T)} \left<
\dfrac{|v_{x_1}|}{v_F} D_1\dfrac{1-R_1|a|^2}{|1-R_1 a^2|^2}
\right>,
\label{btk}
\end{equation}
while a non-local correction to (\ref{btk}) is small in the parameter
${\cal A}_{2}/L^2$ and will be omitted here. The conductance $G_{22}(T)$
of the second interface is defined analogously.

In the temperature interval $ e^{-\Delta / T} \ll 1$
we obtain following expressions for the conductances
\begin{gather}
G_{11}(T)=G_{11}(0) +
G_{N_{11}}\sqrt{2\pi}\sqrt{\dfrac{\Delta}{T}}e^{-\Delta/T},
\label{Gloc1}
\\
G_{12}(T)=
\begin{cases}
2 G_{N_{12}} e^{-\Delta / T}, & \mathcal{D}_1\mathcal{D}_2 \ll e^{-\Delta / T}, \\
G_{12}(0), & T \ll \dfrac{\Delta}{ \ln(1/[\mathcal{D}_1\mathcal{D}_2])},  \\
\end{cases}
\label{Gnonloc1}
\end{gather}
where
\begin{equation}
\frac{G_{12}(0)}{G_{N_{12}}}= \dfrac{{\cal D}_1 {\cal D}_2
(1-\tanh^2 L \Delta/v_F ) }{ [1+ {\cal R}_1 {\cal R}_2 + ({\cal
R}_1 + {\cal R}_2)\tanh L \Delta /v_F]^2}, \label{nonlcon}
\end{equation}
$G_{N_{11}}=\dfrac{2\mathcal{N}_1}{R_q } \left<
\dfrac{|v_{x_1}|}{v_F} D_1(p_{x_1}) \right>$ is the normal state
(Landauer) conductance and $G_{11}(0)=\dfrac{4 {\cal N}_1}{R_q}
\left< \dfrac{|v_{x_1}|}{v_F} D^2_1(p_{x_1}) \right>$ is the subgap
(BTK) conductance of the first NS interface.

Turning now to the non-local resistance
\begin{equation}
R_{12}(T)=\dfrac{G_{12}(T)}{G_{11}(T)G_{22}(T)-G_{12}(T)G_{21}(T)},
\label{Rnonloc}
\end{equation}
we substitute the expressions \eqref{Gnonloc1} and \eqref{Gloc1} into
Eq. \eqref{Rnonloc} and, neglecting the small cross-conductance term
$G_{12}G_{21}$ in the denominator, we obtain
\begin{multline}
R_{12}(T)=\dfrac{2R_{N_{12}} e^{-\Delta/T}}{
\left[G_{11}(0)/G_{N_{11}}+\sqrt{2\pi\Delta /T}
e^{-\Delta/T}\right]}
\times\\\times
\dfrac{1}{ \left[G_{22}(0)/G_{N_{22}}+\sqrt{2\pi\Delta
/T} e^{-\Delta/T}\right]}, \label{Rnonloc2}
\end{multline}
where $R_{N_{12}}=G_{N_{12}}/(G_{N_{11}} G_{N_{22}})$ is the
cross-resistance in the normal state. Eq. \eqref{Rnonloc2} applies
within the interval  $D_1 D_2 \ll  e^{-\Delta / T} \ll 1$  and represents the
central result of this paper.

We observe that the non-local
resistance \eqref{Rnonloc2} first increases with decreasing $T$
reaching its maximum at $T=T^*$ and then decreases with $T$ at lower
temperatures.  For the temperature
$T^*$  with the logarithmic accuracy we obtain
\begin{equation}
T^*\simeq
\dfrac{\Delta}{\ln\sqrt{\dfrac{G_{N_{11}}G_{N_{22}}}{G_{11}(0)G_{22}(0)}}},
\label{Tm}
\end{equation}
or simply $T^*\simeq \Delta /\ln (1/D)$ for symmetric structures
with $D_{1,2}=D$. This result matches qualitatively with that obtained
in Ref. \cite{GZ07} within a different model.

\begin{figure}
\centerline{\includegraphics[width=75mm]{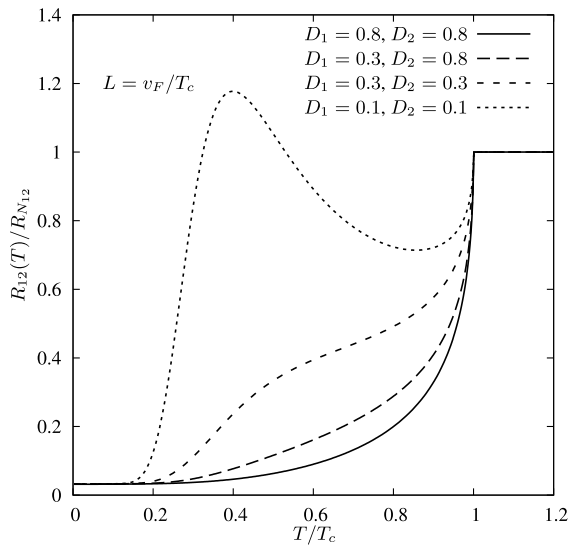}}
\caption{Non-local resistance $R_{12}$ of an NSN device
as a function of temperature for $L=v_F/T_c$. Provided both interface
transmissions are
sufficiently low the non-local resistance exhibits a well pronounced peak.
This peak disappears at higher barrier transmissions.} \label{figr1}
\end{figure}

The temperature dependence of the non-local resistance $R_{12}(T)$
for our NSN device is depicted in Fig. 2 for different values of
the interface transmissions. In the limit of low transmissions the
resistance $R_{12}(T)$ decreases with $T$ right below the
superconducting critical temperature $T_c$ but then turns upwards
and exhibits a well pronounced peak. At lower $T$ the resistance
$R_{12}(T)$ decreases sharply and eventually tends to
$R_{12}(0)\equiv G_{12}(0)/[G_{11}(0)G_{22}(0)]$ in the limit $T \to 0$. With
increasing interface transmissions $D_{1,2}$ the peak gets less
pronounced and eventually disappears, in which case the non-local
resistance $R_{12}(T)$ decreases monotonously with temperature.
The same feature can also be observed in Eq. (\ref{Rnonloc2}).

\begin{figure}
\centerline{\includegraphics[width=75mm]{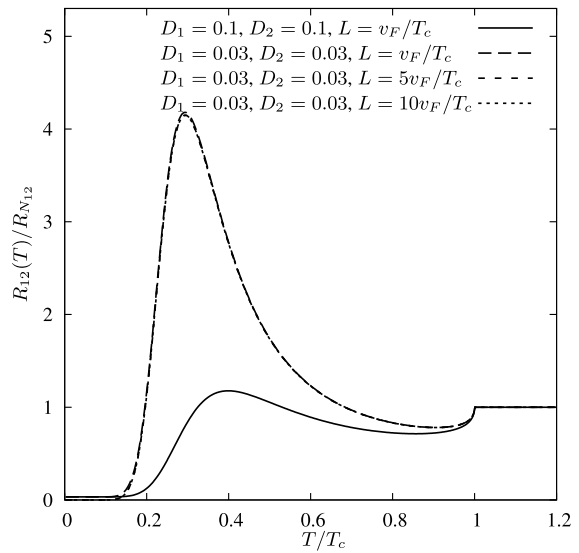}} \caption{ The same
as in Fig. 2 for different values of the distance $L$ between two
N-terminals (measured in units of $\xi$). Three curves
corresponding to the same barrier transmissions but different
values of $L$ practically coincide, i.e. in the vicinity of the
peak $R_{12}(T)$ in the superconducting state scales with $L$
exactly as $R_{N_{12}}$.} \label{figr2}
\end{figure}

As it is clear from Fig. 3 both the height and the form of the
non-local resistance peak scale with the distance $L$ exactly as
the normal state value $R_{N_{12}}$. For instance, from Eq.
(\ref{Rnonloc2}) we easily determine the height of the resistance
peak $R_{12}(T^*)$ which reads
\begin{equation}
\frac{R_{12}(T^*)}{R_{N_{12}}} =\dfrac{\sqrt{\dfrac{2T^*}{\pi\Delta}}}{
\left[
\sqrt{\dfrac{G_{11}(0)}{G_{N_{11}}}}+\sqrt{\dfrac{G_{22}(0)}{G_{N_{22}}}}
\right]^2}
\label{height}
\end{equation}
We observe that  the right-hand side of Eq. (\ref{height}) is expressed
only in terms of local conductances and, hence, is independent of $L$.
This is an important result which might account for experimental
observations \cite{Venkat} of a  much weaker $L$-dependence
of $R_{12}(T^*)$ as compared to the zero temperature value
$R_{12}(0)$. For instance, within our model of ballistic
electrodes for $L \gg \xi_0$ we have $R_{12}(0)\propto \exp (-2L\Delta(0)/v_F )$
while  $R_{12}(T^*) \propto 1/L^2$. Even weaker length dependence of
$R_{12}(T^*)\propto R_{N_{12}} \propto 1/L$ is expected in the diffusive
limit. It would be interesting to scale the data \cite{Venkat}
for the resistance peak at different lengths $L$ with the
corresponding normal state resistance $R_{N_{12}}(L)$.

Although our theory correctly accounts for some key features of
the experimental data it is important to bear in mind that the
model employed here deals with ballistic electrodes connected via
metallic constrictions whereas in experiments
\cite{Beckmann,Venkat} the electrodes were most likely diffusive.
In addition, we disregarded any relaxation mechanisms for
non-equilibrium quasiparticles inside the superconductor (except
for their escape into the normal terminals) while such mechanisms
(caused, e.g., by electron-phonon and electron-electron
interactions) are obviously present in experiments being
responsible for a finite charge imbalance length. Hence, one can
also expect certain differences. For instance, within our model
the non-local resistance peak occurs only at small interface
transmissions while the authors \cite{Beckmann,Venkat} observed
such a peak at moderately high transmissions of NS interfaces.
This might be an indication to relatively more pronounced charge
imbalance effects.

In summary, we have developed a quantitative theory describing the
peak in the temperature dependence of the non-local resistance
recently observed in three-terminal NSN devices
\cite{Beckmann,Venkat} at sufficiently high $T$. This peak emerges
as a result of a trade-off between quasiparticle/charge imbalance
and subgap (Andreev) contributions to local and non-local
conductances of the device. Both the height and the shape of the
peak scale with the normal state resistance $R_{N_{12}}$ thus
demonstrating much weaker (power law) dependence on the
superconductor thickness $L$ as compared to the zero-temperature
resistance $R_{12}(0)$ which decays (roughly) exponentially with
increasing $L$.

We acknowledge stimulating discussions with D. Beckmann, V.
Chandrasekhar and especially with D.S. Golubev. This work was
supported in part by RFBR grant 06-02-17459.

\end{document}